\documentclass[aps,twocolumn,pra,superscriptaddress,amsmath,amssymb]{revtex4-2}
\usepackage{dcolumn}
\usepackage{bm}
\usepackage{amsmath}
\usepackage{txfonts}
\usepackage[T1]{fontenc}
\usepackage{xspace}
\usepackage{ulem}
\usepackage{comment}
\usepackage{braket}
\ifx\pdfoutput\undefined
\usepackage[dvipdfmx]{graphicx}
\usepackage[dvipdfmx]{hyperref}
\usepackage[dvipdfmx]{xcolor}
\else
\usepackage{graphicx}
\usepackage{hyperref}
\usepackage{xcolor}
\fi

\newcommand{\cred}[1]{\textcolor{black}{#1}}

\begin{document}
\let\emph\textit

\title{
  Triangular and dice quasicrystals modulated by generic 1D aperiodic sequences
}
\author{Toranosuke Matsubara}
\author{Akihisa Koga}

\affiliation{
  Department of Physics, Institute of Science Tokyo,
  Meguro, Tokyo 152-8551, Japan
}

\author{Tomonari Dotera}

\affiliation{
  Department of Physics, Kindai University,
  Higashi-Osaka, Osaka 577-8502, Japan
}

\date{\today}
\begin{abstract}
We present a method for generating hexagonal aperiodic tilings that are topologically equivalent to the triangular and dice lattices. This approach incorporates aperiodic sequences into the spacing between three sets of grids for the triangular lattice, resulting in ``modulated triangular lattices''. Subsequently, by replacing the triangles with rhombuses, parallelograms, or hexagons, modulated dice or honeycomb lattices are constructed. Using generalized Fibonacci, Thue-Morse, and tribonacci sequences, we demonstrate several examples of hexagonal aperiodic tilings. Structural analysis confirms that their diffraction patterns reflect the properties of the one-dimensional aperiodic sequences, namely pure point (Bragg peaks) or singular continuous. Our method establishes a general framework for constructing a broad range of hexagonal aperiodic systems, advancing aperiodic-crystal research into higher dimensions that were previously focused on one-dimensional aperiodic sequences.

\end{abstract}
\maketitle

\section{Introduction}

Quasicrystals have been a subject of active research
since the discovery of the Al-Mn alloy~\cite{shechtman_1984,Levine}.
Unlike crystals, quasicrystals lack spatial periodicity but exhibit unique long-range ordered structures. 
It is well known that typical quasicrystalline structures can be modeled using quasiperiodic tilings, 
such as the decagonal Penrose~\cite{Penrose,Mackay_1982},
octagonal Ammann-Beenker~\cite{Beenker,Ammann},
dodecagonal Stampfli~\cite{Stampfli},
and three-dimensional Penrose tilings~\cite{kramer}.
A key challenge in studying quasicrystalline structures lies in understanding the similarities and differences between quasicrystalline and crystalline structures.

One approach involves analyzing specific series of periodic crystals and studying their limits as quasicrystals. 
These periodic crystals, referred to as ``periodic approximants''~\cite{Goldman}, converge toward the quasicrystal as the degree of approximation increases. 
Systematic experimental studies in this field have revealed intriguing phenomena in quasicrystals, 
such as the quantum critical state in Au-Al-Yb~\cite{Ishimasa_2011,Deguchi}, 
the superconducting state in Al-Mn-Zn~\cite{Ramachandrarao_1985,Kamiya}, 
and the magnetically ordered state in \cred{Au-Ga-$R$ ($R$ = Gd, Tb, and Dy)}~\cite{Tamura_2021, Takeuchi_Farid}. 
Furthermore, the concept of ``aperiodic approximants'' has been introduced as an alternative framework~\cite{Nakakura_2019,Matsubara_honey,Nakakura_2023}, 
where specific series of quasicrystals are analyzed, and their limits as periodic crystals are explored. 
The structural similarities between periodic crystals and quasicrystals, as demonstrated by these two series, continue to provide valuable insights into their shared and unique properties.

Another important approach involves the continuous modulation of the periodic lattice~\cite{WJJ1981}. 
A straightforward extension to two dimensions is the square Fibonacci tiling~\cite{stinchcombe_1990, lifshitz_2002, Even_2008}, 
which is constructed by incorporating Fibonacci sequences into the spacing between the orthogonal grid lines of a square lattice. Since the length ratio can be continuously varied, the square Fibonacci tiling is topologically equivalent to the square lattice. This continuous modulation thus serves as a method for systematically exploring the structural similarities between crystals and quasicrystals. In contrast to the well-studied tetragonal case, hexagonal aperiodic structures remain largely unexplored, 
although similar hexagonal tilings have been generated using a generalized de Bruijn dual grid method~\cite{Bruijn_1981,Bruijn_1986,Socolar_1985,Gahler_1986,Ho_1986,Rabson_1988, Rabson_1989, Lifshitz_2005,Sam_8GMT}. Consequently, developing a method for introducing aperiodic modulations into triangular lattices is highly desirable.

In this study, we propose a method for generating hexagonal aperiodic tilings that are topologically equivalent to the triangular lattice. By employing aperiodic sequences in the line spacing of the three grids in the triangular lattice, we obtain modulated lattices. These tilings exhibit a rich variety, in contrast to those derived from the square lattice. For example, by replacing triangles with rhombuses and parallelograms (\cred{hexagons}), we obtain a decorated lattice that is topologically equivalent to the dice (\cred{honeycomb}) lattice. We demonstrate several aperiodic hexagonal tilings, utilizing generalized Fibonacci, Thue-Morse, and tribonacci sequences. The substitution rules are derived by comparing the tiling with its enlarged version, scaled by the characteristic ratio of the sequence. Additionally, the structure factors of these tilings are analyzed.

This paper is organized as follows. In Sec.~\ref{sec: method}, we describe the continuous modulation of a simple periodic lattice using a Fibonacci sequence. In Sec.~\ref{sec: tiling}, we extend this method to the triangular lattice with a generic aperiodic sequence and generate two-dimensional hexagonal tilings. We then demonstrate hexagonal metallic-mean, Thue-Morse, and tribonacci tilings. A summary is provided in the final section, and additional details are included in Appendices.

\section{Method}\label{sec: method}

We begin by explaining the continuous modulation
of the periodic lattice using an aperiodic sequence.
Here, we consider an aperiodic sequence composed of specific letters.
A simple example is the Fibonacci sequence, which is composed of the letters L and S.
This is generated by iteratively applying
the substitution rules L $\rightarrow$ LS and S $\rightarrow$ L.
The fractions of these letters are given as
$f_{\rm L}=\tau_1^{-1}$ and $f_{\rm S}=\tau_1^{-2}$ with the golden mean $\tau_1[=(1+\sqrt{5})/2]$.

\begin{figure}[htb]
  \begin{center}
    \includegraphics[width=\linewidth]{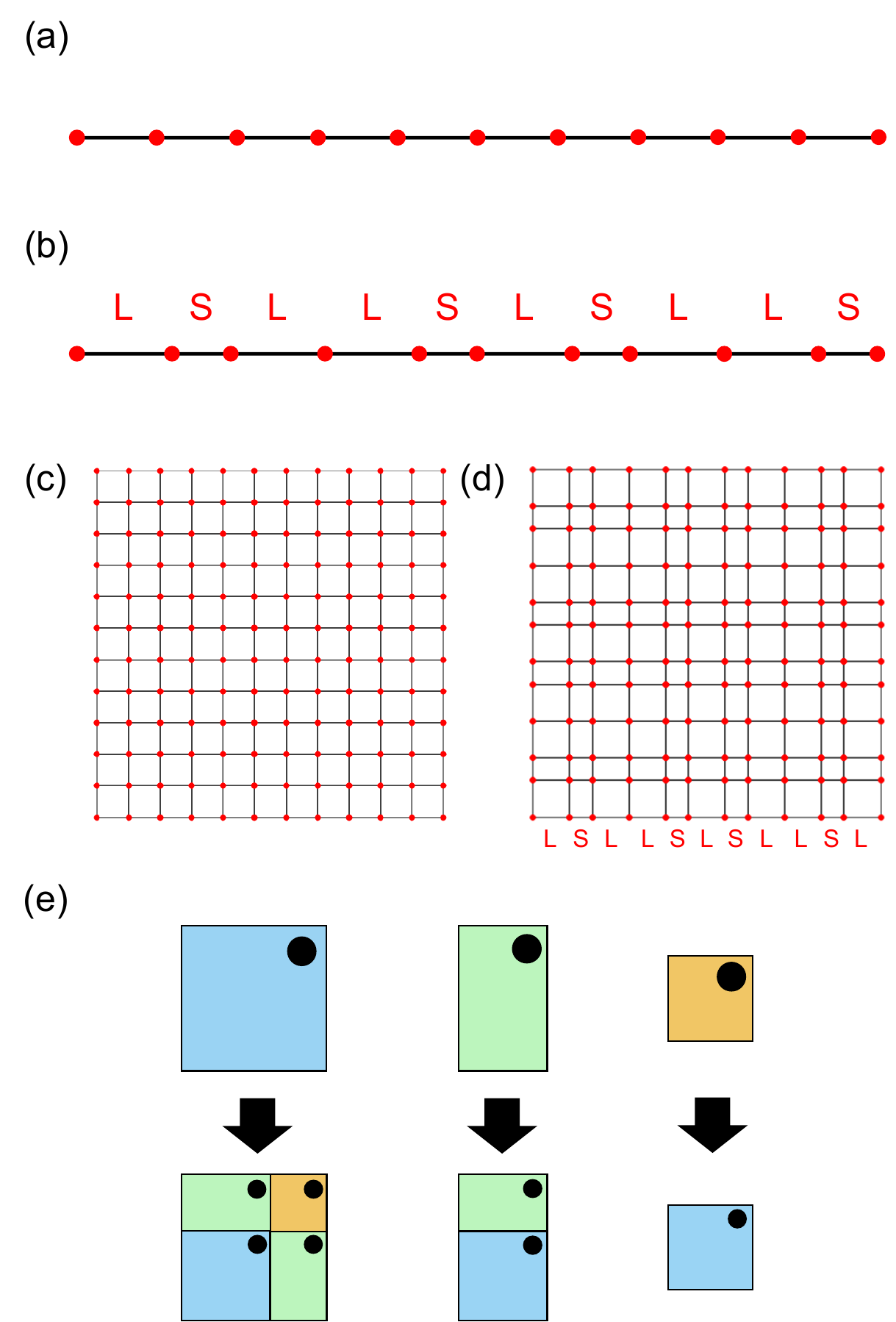}
    \caption{(a) One-dimensional regular lattice.
    (b) One-dimensional lattice modulated by the Fibonacci sequence.
    (c) Square lattice. (d) Square Fibonacci lattice and (e) its substitution rule.
    }
    \label{fig: 1D2D}
  \end{center}
\end{figure}
We here consider a one-dimensional lattice with a lattice constant $a$,
which is schematically shown in Fig.~\ref{fig: 1D2D}(a).
The letters L and S in the Fibonacci sequence are assigned to the bonds
between nearest neighbor vertices, with corresponding lengths defined as $\ell$ and $s$.
This leads to the one-dimensional Fibonacci lattice,
which is shown in Fig.~\ref{fig: 1D2D}(b).
When $\ell f_{\rm L} + s f_{\rm S} = a$,
the density of the vertices remains unchanged
despite the introduction of the Fibonacci sequence.
Consequently, we obtain the modulated one-dimensional chain
for arbitrary $\ell/s$.
Namely, when $\ell=s$, it is reduced to the original one-dimensional lattice,
as shown in Fig.~\ref{fig: 1D2D}(a).
It is important to note that an exact self-similar structure emerges spatially
when the length ratio is fixed as the characteristic ratio of the Fibonacci sequence $\ell /s = \tau_1$.
This can be observed by comparing the one-dimensional Fibonacci lattice
with its enlarged lattice by $\tau_1$.

We now extend our discussions to two dimensions,
starting with a trivial example of a square lattice
with lattice constant $a$, as shown in Fig.~\ref{fig: 1D2D}(c).
We employ the Fibonacci sequence in the interline spacing for each orthogonal grid
shown in Fig.~\ref{fig: 1D2D}(d).
Then, we derive the square Fibonacci tiling as a modulated square lattice~\cite{stinchcombe_1990, lifshitz_2002, Even_2008},
where two lengths are given by $\ell$ and $s$.
This tiling is composed of large squares (LS), rectangles (R), and small squares (SS).
Fixing the length ratio as $\ell/s=\tau_1$ and 
comparing the tiling with its enlarged tiling by $\tau_1$,
we observe a self-similar structure in the tiling.
This leads to the substitution rule for directed tiles~\cite{lifshitz_2002},
as shown in Fig.~\ref{fig: 1D2D}(e).
The numbers of the tiles increases under the substitution rule,
which is represented as
${\bf v}^{(n+1)}=M{\bf v}^{(n)}$,
where
${\bf v}^{(n)}=(N_{\rm LS}^{(n)}, N_{\rm SS}^{(n)},N_{\rm R}^{(n)})^t$
and $N_j^{(n)}$ is the number of $j$th tile at iteration $n$, and
the substitution matrix $M$ is given as
\begin{equation}
  M=\left(\begin{array}{lll}
    1 & 1 & 1\\
    1 & 0 & 0\\
    2 & 0 & 1 
  \end{array}\right).
\end{equation}
The maximum eigenvalue of $M$ is $\tau^2_1$,
meaning that the tiling has the self-similar golden-mean structure.
The corresponding eigenvector is given
by
$(\tau_1^{-2}, \tau_1^{-4}, 2\tau_1^{-3})^t$.
Then, the fractions for the tiles are exactly obtained.
Since the square Fibonacci tiling can be regarded
as the modulated square lattice,
these fractions remain constant for arbitrary length ratio $\ell/s$.
Consequently, the self-similar structure found in the special case
should be useful for analytically evaluating the tile fractions
in the modulated lattice.
In the following, we will employ the aperiodic sequence
in the interline spacing of grids of the triangular lattice to generate the hexagonal tilings.

\section{Aperiodic hexagonal tilings}\label{sec: tiling}

\subsection{Golden-mean tiling}

\begin{figure*}[htb]
  \begin{center}
    \includegraphics[width=\linewidth]{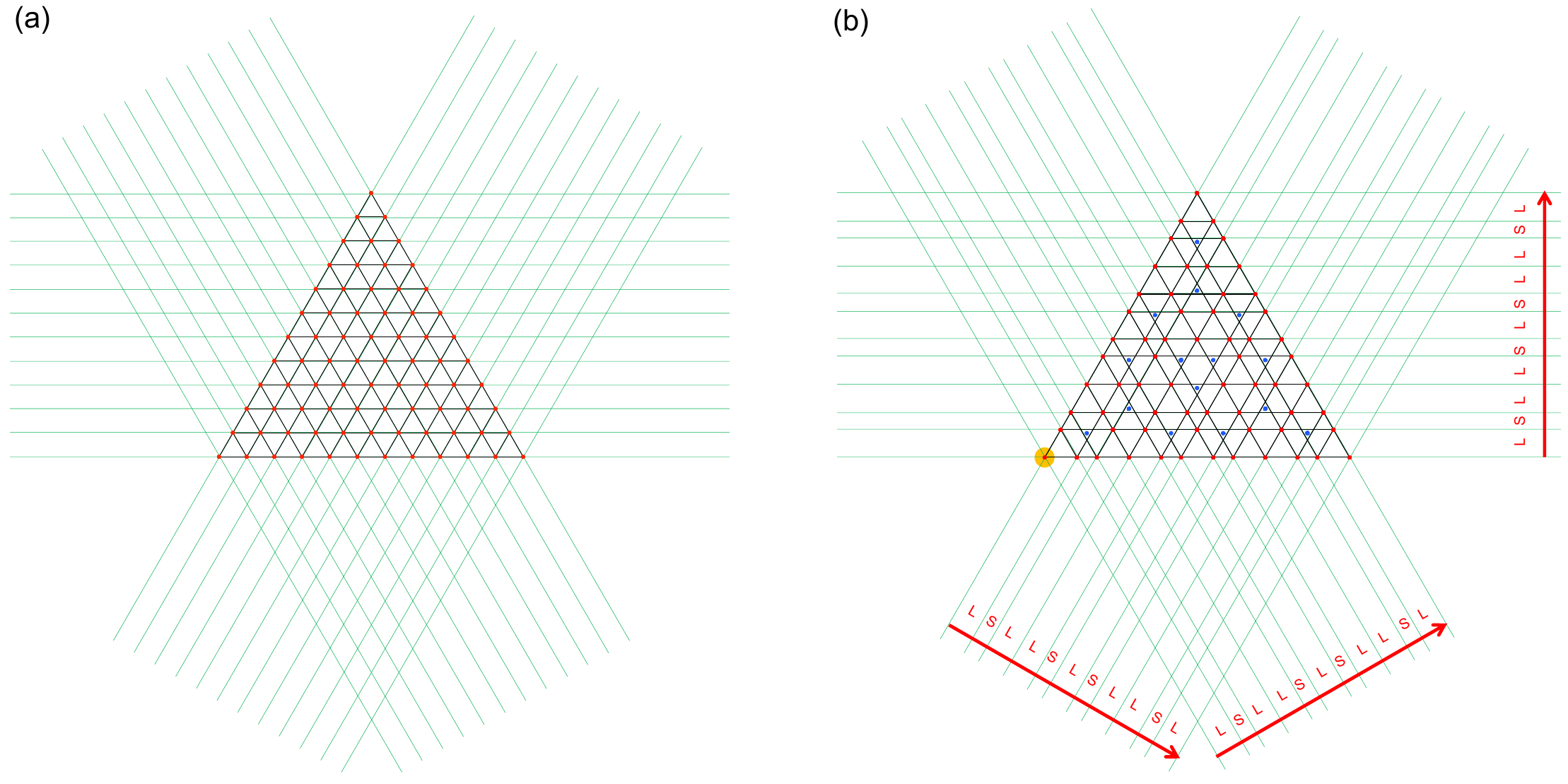}
    \caption{(a) Regular triangular lattice. Red circles represent the intersections of three grids.
    (b) Three sets of grids modulated by the Fibonacci sequence, which 
      is represented by L and S.
      \cred{Originating from a reference point marked in orange, grids with aperiodic interline spacings extend in three distinct directions indicated by arrows.
      }
      Blue circles represent the center of three intersections
      split by the introduction of the aperiodic sequence
      and red circles represent the unsplit intersections (see text).
    }
    \label{fig: triangle1}
  \end{center}
\end{figure*}

\begin{figure}[htb]
  \begin{center}
    \includegraphics[width=\linewidth]{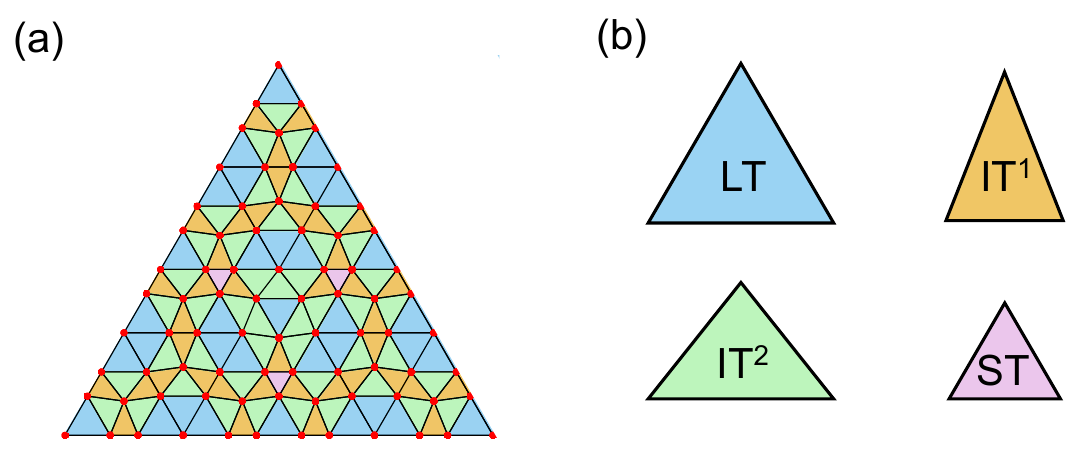}
    \caption{(a) Triangular lattice modulated by the Fibonacci sequence
    and
    (b) \cred{four triangles} LT, IT$^1$, IT$^2$, and ST.
    }
    \label{fig: triangle2}
  \end{center}
\end{figure}

\begin{figure}[htb]
  \begin{center}
    \includegraphics[width=0.8\linewidth]{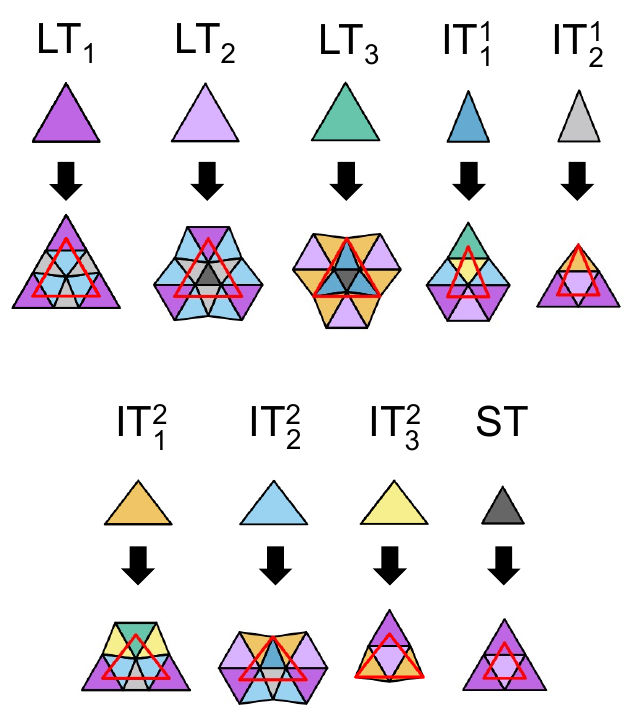}
    \caption{
      Substitution rules for the Fibonacci-modulated triangular lattice which is shown in Fig.~\ref{fig: triangle2}.
      Distinct colored tiles represent the \cred{nine prototiles}.
    }
    \label{fig: sub-gmt}
  \end{center}
\end{figure}

Let us consider a triangular lattice modulated by the Fibonacci sequence.
The regular triangular lattice, characterized by a lattice constant $a$,
can be viewed as superimposing three sets of grids, each rotated by 120$^\circ$, with a common interline spacing $(\sqrt{3}/2) a$, as shown in Fig.~\ref{fig: triangle1}(a).
\cred{The Fibonacci sequences are employed
to determine the interline spacings of each grid,
with the two spacings given by $(\sqrt{3}/2)\ell$ and $(\sqrt{3}/2)s$.
In Fig.~\ref{fig: triangle1}(b), originating from a reference point marked in orange, grids with aperiodic interline spacings extend in three distinct directions indicated by arrows.}
Unlike the square Fibonacci tiling, a different problem arises here due to non-orthogonality of the three grids.
In a regular \cred{triangular} lattice, three grids always meet at intersections.
However, in our case, three grids do not always meet at intersections;
instead, they become split into three intersections formed by pairs of grids.
\cred{The resulting aperiodic tiling} is composed of three sizes of \cred{regular} triangles and one type of trapezoids.

To eliminate trapezoids and construct a triangular lattice, we merge each three closest points into a single point at their center displayed by a blue circle in Fig.~\ref{fig: triangle1}(b), resulting in the modulated triangular lattice shown in Fig.~\ref{fig: triangle2}(a). Although the lattice is mostly composed of zig-zag lines and thus appears to be deformed, it should be noticed that the number of vertices and its topology remain the same as those in the regular triangular lattice.
Accordingly, there exist four kinds of \cred{triangles} in the tiling as shown in Fig.~\ref{fig: triangle2}(b):
\cred{
small or large regular triangles denoted by ST or LT with its edge length $s$ or $\ell$, respectively;
}
\cred{small or large isosceles triangles} denoted by IT$^1$  or IT$^2$ with isosceles edge length $m$ and the other edge length, short ($s$) or long ($\ell$), respectively, where $m=s\sqrt{1+r+r^2}/\sqrt{3}$ and $r=\ell/s$.

Fixing the length ratio as $\ell/s=\tau_1$ and
comparing the aperiodic tiling with its enlarged one by $\tau_1$,
we find the self-similar structure.
Figure~\ref{fig: sub-gmt} shows the substitution rules with nine triangles,
LT$_1$, LT$_2$, LT$_3$, IT$^1_1$, IT$^1_2$, IT$^2_1$, IT$^2_2$ IT$^2_3$, and ST.
To calculate the frequencies of tiles, the substitution matrix can be derived.
However, its matrix elements remain complex.
This originates from the fact that two distinct lengths $\ell$ and $s$ yield the other length $m$,
resulting in the \cred{small and large isosceles triangles} with intricate angles even when $\ell/s=\tau_1$.

\begin{figure}[htb]
  \begin{center}
    \includegraphics[width=\linewidth]{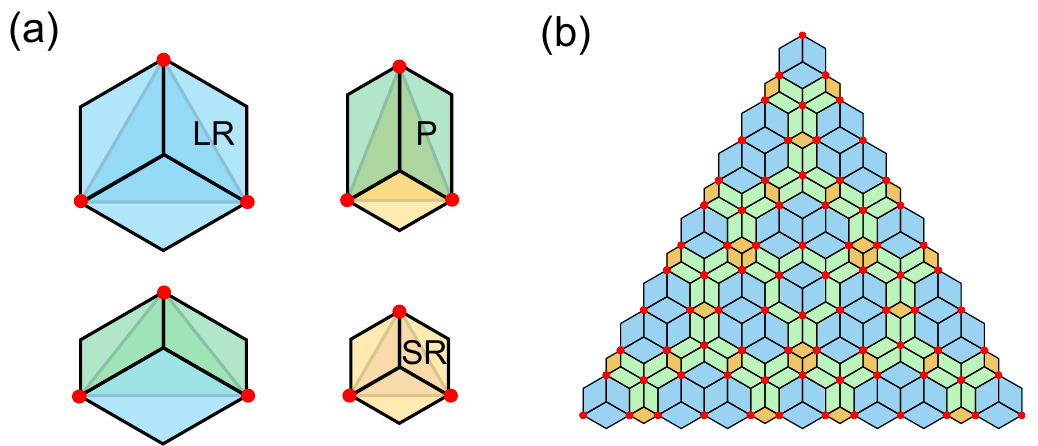}
    \caption{(a) Decoration rules for transforming four types of triangles into large and small rhombuses, LR and SR, and parallelogram P.
 (b) Fibonacci-modulated dice lattice.}
    \label{fig: Dice}
  \end{center}
\end{figure}

To overcome the above complexity, we 
substitute four kinds of triangles with polygons having angles that are multiples of 60$^\circ$, leading to quadrilateral tilings composed of large rhombus (LR), parallelogram (P), and small rhombus (SR) as shown in Fig.~\ref{fig: Dice}.
\begin{figure}[htb]
  \begin{center}
    \includegraphics[width=0.9\linewidth]{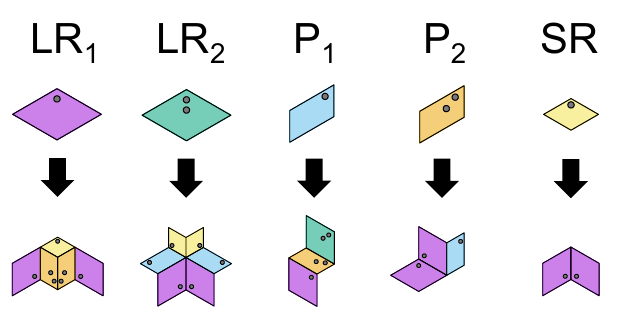}
    \caption{
      Substitution rules for the Fibonacci-modulated dice lattice.
      Distinct colored rhombuses and parallelograms represent
      the \cred{five prototiles}.
      }
    \label{fig: sub-dc}
  \end{center}
\end{figure}
Here, their vertices with an acute angle 60$^\circ$ are located at the vertices on the modulated triangular lattice.
Since the obtained tiling structure is topologically equivalent to the dice lattice,
it can be regarded as
the dice lattice modulated by the Fibonacci sequence.
By comparing with the enlarged tiling by $\tau_1$,
we obtain the substitution rules for five \cred{prototiles},
as shown in Fig.~\ref{fig: sub-dc}.
The substitution matrix for the tiles, LR$_1$, LR$_2$, P$_1$, P$_2$, and SR
is represented by rational numbers as
\begin{equation}
  M=\left(\begin{array}{lllllllll}
  1 & 1 & \frac{1}{2} & 1 & 1 \\
  0 & 0 & \frac{1}{2} & 0 & 0 \\
  0 & 2 & 0           & 1 & 0 \\
  2 & 0 & 1           & 0 & 0 \\
  1 & 1 & 0           & 0 & 0 
  \end{array}\right).
\end{equation}
The maximum eigenvalue of $M$ is $\tau^2_1$,
meaning that the tiling has the self-similar golden-mean structure.
The corresponding eigenvector is given
by $(3\tau_1^{-3} /2, \tau_1^{-6}/2, \tau_1^{-4}, \tau_1^{-3} + \tau^{-5}_1,\tau_1^{-4})^t$,
which represents the fractions of tiles, LR$_1$, LR$_2$, P$_1$, P$_2$, and SR, respectively.
Therefore, the frequencies of large and small rhombuses, and parallelograms are given by
\begin{align}
  f_{\rm LR} &= f_{\rm LR1}+f_{\rm LR2}=\frac{1}{\tau_1^2}\approx 0.382,\\
  f_{\rm P} &= f_{\rm P1}+f_{\rm P2}=\frac{2}{\tau_1^3} \approx 0.472,\\
  f_{\rm SR} &= \frac{1}{\tau^4_1} \approx 0.146.
\end{align}

\begin{figure}[htb]
  \begin{center}
    \includegraphics[width=\linewidth]{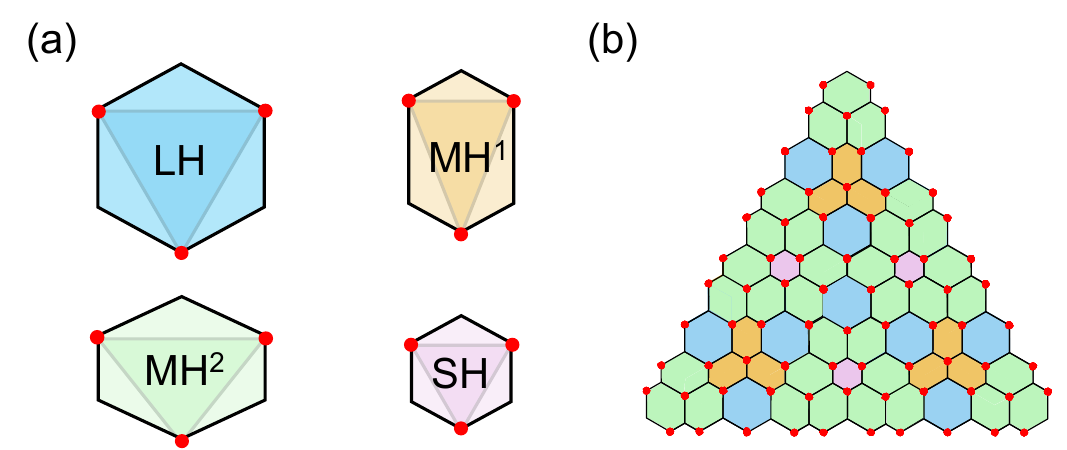}
    \caption{
    (a) The replacement of downward triangles, 
    where LH, MH$^1$, MH$^2$, SH tiles are introduced instead of triangles.
      (b) Fibonacci-modulated honeycomb lattice formed by the rule (a).
    }
    \label{fig: honey_down}
  \end{center}
\end{figure}

\begin{figure}[htb]
  \begin{center}
    \includegraphics[width=0.7\linewidth]{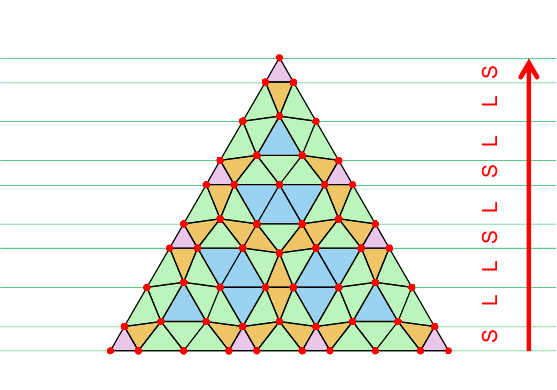}
    \caption{
      A different class of the Fibonacci-modulated triangular lattice,
      where the Fibonacci sequence starting from the second term is introduced.
      As an example, the vertically aligned sequence in the grids is illustrated.
    }
    \label{fig: second}
  \end{center}
\end{figure}

To add an important kind of tiles, another decorated tiling can be obtained
by replacing triangles with hexagons having inner angles of 120$^\circ$.
In the triangular lattice, upward and downward triangles are arranged alternately. 
By replacing downward triangles in the modulated triangular lattice
with large hexagons (LH), medium hexagons (MH$^1$ and MH$^2$),
and small hexagons (SH),
a modulated honeycomb lattice is formed, as shown in Fig.~\ref{fig: honey_down}.
While one may presume that this modulated honeycomb lattice is simpler than the dice lattice, it is, in fact, somewhat more complex due to the greater variety of 
\cred{tiles}. Moreover, the modulated honeycomb lattice can be easily derived by excluding appropriate vertices from the \cred{modulated} dice lattice. In the following sections, we demonstrate the modulated triangular and dice lattices as examples.

\begin{figure*}[htb]
  \begin{center}
    \includegraphics[width=0.9\linewidth]{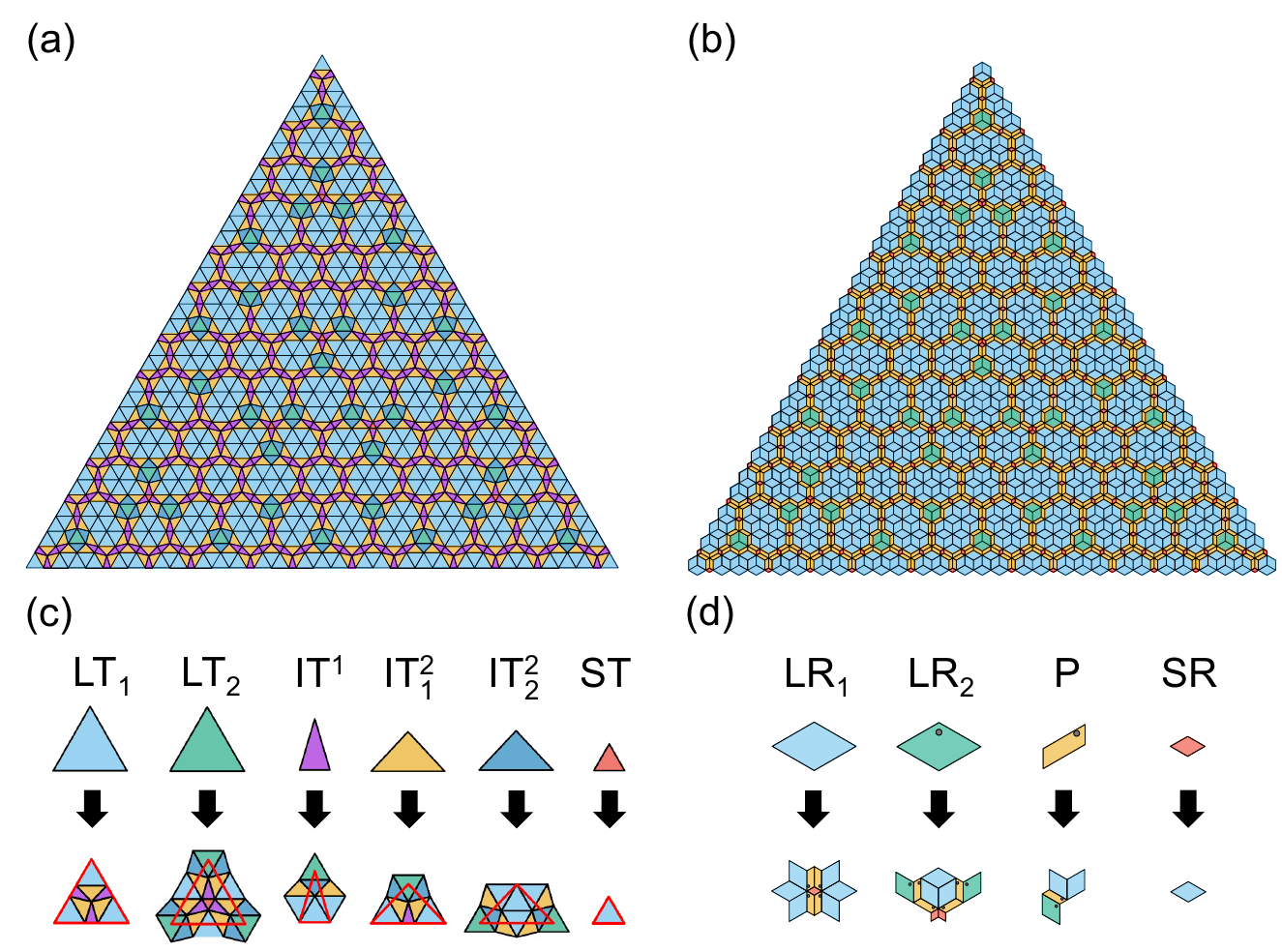}
    \caption{Silver-mean modulated lattices and their substitution rules. (a) Triangular lattice.
    (b) Dice lattice. (c) Substitution rules for (c) triangular and (d) dice lattices. 
      Colored tiles correspond to these substitution rules (c) and (d). 
    }
    \label{fig: SMT}
  \end{center}
\end{figure*}

We emphasize that when the Fibonacci sequence is assigned to the interline spacing
of each grid in the triangular lattice, there exist degrees of freedom in the arrangement.
Figure~\ref{fig: second} shows a modulated triangular lattice in which the Fibonacci sequence with nine letters is introduced starting from the second term. 
It is evident that this tiling differs from that shown in Fig.~\ref{fig: triangle2}(a)
since the density of ST tiles is different from each other.
This flexibility arises from the non-orthogonality of the triangular lattice grids, a feature absent in the square Fibonacci lattice, providing a greater variety of tiling structures. 
These different tilings belong to different ``local isomorphism classes~\cite{Levine_1986,Socolar_1986}''.
In Fig.~\ref{fig: triangle1}(b), the grid spacing is described by
\begin{eqnarray}
\frac{x_N}{(\sqrt{3}/2)s}&=&N-1+\frac{1}{\tau_1}\left\lfloor\frac{N}{\tau_1}\right\rfloor, 
\, (N=1, 2, \cdots, 12),
\end{eqnarray}
while in Fig.~\ref{fig: second} the grid spacing is described by
\begin{eqnarray}
\frac{x_N}{(\sqrt{3}/2)s}&=&N-\tau_1+\frac{1}{\tau_1}\left\lfloor\frac{N+1}{\tau_1}\right\rfloor,
\, (N=1, 2, \cdots, 10),
\end{eqnarray}
where the step function $\left\lfloor x \right\rfloor$ is the integer less than or equal to $x$ and closest to $x$,
$x_N$ gives the position of the $N$th line of grid,
and this formula corresponds to the case $\ell /s = \tau_1$.
These phase differences produces different tilings,
 and are known to affect the third order term of Landau free energies as $\cos(\phi_1+\phi_2+\phi_3)$~\cite{Bak_1985}, 
 where $\phi_i$ is the phase of the density wave description in each direction.
Further details are discussed in Appendix~\ref{app}.
We also note that $H_{00}$ and $H_{\frac{1}{2} \frac{1}{2}}$
hexagonal golden-mean tilings proposed recently are also \cred{parts} of this family, as discussed in Appendix~\ref{sec: honey}.

In the above discussion, we have focused on the Fibonacci sequences
to generate golden-mean modulated quasiperiodic tilings.
It is important to note that one can also introduce more generalized sequences
such as generalized Fibonacci, Thue-Morse, and tribonacci sequences.
In the following,
we will present several examples for the hexagonal aperiodic tilings.

\begin{figure*}[htb]
  \begin{center}
    \includegraphics[width=0.9\linewidth]{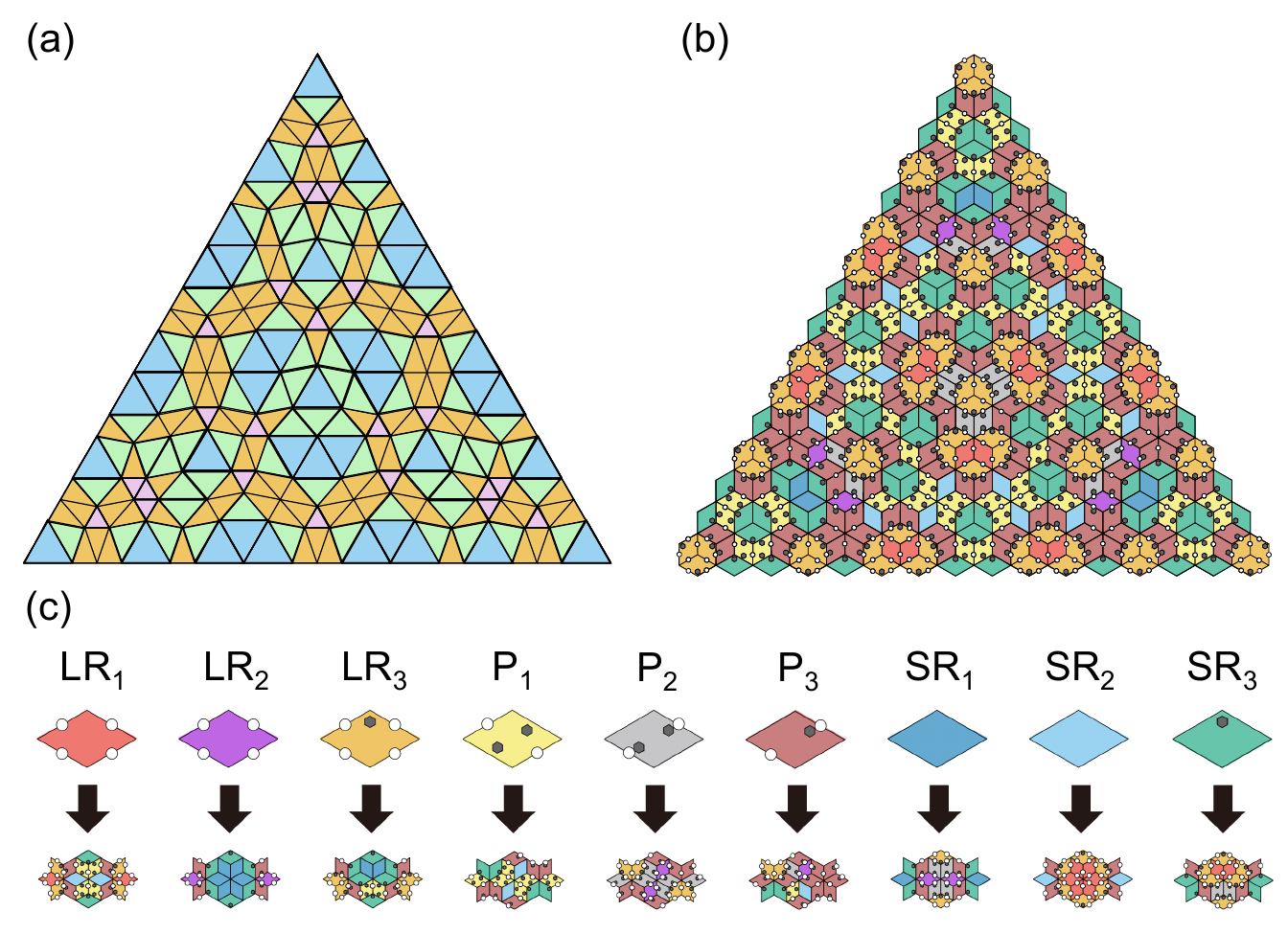}
    \caption{
      (a) Thue-Morse modulated triangular lattice with $\ell /s = 2$.
      Blue, orange, green and pink triangles represent
      LT, IT$^1$, IT$^2$ and ST, respectively.    
      (b) Thue-Morse modulated dice lattice with $\ell/s=1$, and
      (c) its substitution rules.
      The open circles at the centers of edges represent L bonds (see text).
    }
    \label{fig: TMT}
  \end{center}
\end{figure*}

\subsection{Metallic-mean tilings}
We consider the generalized Fibonacci sequences
which are known to have one-dimensional quasiperiodic structures.
The sequences are shown as
\begin{align}
  k=1 :\,\, & {\rm LSLLSLSLLSLLSLSLLSLSL}\cdots \nonumber,\\
  k=2 :\,\, & {\rm LLSLLSLLLSLLSLLLSLLSL}\cdots \nonumber,\\
  k=3 :\,\, & {\rm LLLSLLLSLLLSLLLLSLLLS}\cdots \nonumber,
\end{align}
which can be generated by
the substitution rule: L $\rightarrow$ LLL$\cdots$S = L$^k$S and
S $\rightarrow$ L, where $k$ is integer.
The numbers of the letters L and S at iteration $n$
($N_{\rm L}^{(n)}$ and $N_{\rm S}^{(n)}$) satisfy 
\begin{equation}
\left(\begin{array}{l}
N_{\rm L}^{(n+1)} \\
N_{\rm S}^{(n+1)}
\end{array}\right)=\left(\begin{array}{ll}
k & 1 \\
1 & 0
\end{array}\right)\left(\begin{array}{l}
N_{\rm L}^{(n)} \\
N_{\rm S}^{(n)}
\end{array}\right),
\end{equation}
where the maximum eigenvalue of the matrix is given by the metallic-mean
$\tau_k = (k + \sqrt{k^2 + 4})/2$.
When $k=1$, this sequence is reduced to \cred{the conventional Fibonacci one
discussed above}.

The generalized Fibonacci sequences are assigned into the spacing
between the grids of the triangular lattice,
where two length scales $\ell$ and $s$, are introduced.
Figure~\ref{fig: SMT}(a)
shows the quasiperiodic silver-mean tilings with $k=2$ and $\ell/s=\tau_2$,
which is topologically equivalent to the triangular lattice.
When one replaces the triangles with rhombuses and parallelograms,
we obtain a distinct quasiperiodic silver-mean tiling,
as shown in \cred{Fig.~\ref{fig: SMT}(b)}.
The self-similar structure spatially appears 
when $\ell/s=\tau_2$.
One can obtain a substitution rule
by comparing the original tiling with the tiling enlarged by $\tau_2$
since this generalized Fibonacci sequence has the self-similar structure.

For the modulated triangular and dice lattices,
the substitution rules are obtained, 
as schematically shown in Figs.~\ref{fig: SMT}(c) and (d).
In the latter case, the substitution matrix $M$ for LR$_1$, LR$_2$, P, and SR tiles
is easily obtained as
\begin{equation}
M=\left(\begin{array}{llll}
  4 & 3 & \frac{3}{2} & 1 \\
  0 & 1 & \frac{1}{2} & 0 \\
  4 & 4 & 1           & 0 \\
  1 & 1 & 0           & 0 
\end{array}\right).
\end{equation}
The maximum eigenvalue of $M$ is given by $\tau^2_2$,
meaning that the tiling has the self-similar silver-mean structure.
The corresponding eigenvector is given
by $(1/2 -\tau_2^{-2}/4, \tau_2^{-2}/4, \tau_2^{-1},1/2\tau_2^{-2})^t$,
which represents the fractions of tiles.
Namely, the fractions of large and small rhombuses, and parallelograms are also given by
\begin{align}
  f_{\rm LR} &= f_{\rm LR1}+f_{\rm LR2}=\frac{1}{2} = 0.5,\\
  f_{\rm P} &= \frac{1}{\tau_2} \approx 0.414,\\
  f_{\rm SR} &= \frac{1}{2\tau^2_2} \approx 0.086.
\end{align}
  We find that the fraction of large (small) rhombus $f_{\rm LR}$ ($f_{\rm SR}$) with $k=2$
  is larger (smaller) than that with $k=1$.
  Furthermore,
  the domains composed of adjacent LR tiles also become larger
  in the modulated dice lattice [see Fig.~\ref{fig: Dice}(b) and Fig.~\ref{fig: SMT}(b)].
  These are consistent with the fact that
  the fraction of the letter L (S)
  in the generalized Fibonacci sequence increases (decreases) with increasing $k$.
Employing the generalized Fibonacci sequences, 
one can construct metallic-mean tilings systematically.
The series of the metallic-mean tilings can be regarded as aperiodic approximants
of the triangular, dice, and honeycomb lattices~\cite{Nakakura_2019, Matsubara_honey,Nakakura_2023}.

\subsection{Thue-Morse tiling}

Here, we consider the Thue-Morse sequence as
\begin{equation}
{\rm LSSLSLLSSLLSLSSLSLLSLSSLLSSLSLLS} \cdots.\nonumber
\end{equation}
This sequence can be generated by s substitution rule
as
L $\rightarrow$ LS and S $\rightarrow$ SL.
The numbers of the letters L and S at iteration $n$
($N_{\rm L}^{(n)}$ and $N_{\rm S}^{(n)}$) satisfy 
\begin{equation}
\left(\begin{array}{l}
N_{\rm L}^{(n+1)} \\
N_{\rm S}^{(n+1)}
\end{array}\right)=\left(\begin{array}{ll}
1 & 1 \\
1 & 1
\end{array}\right)\left(\begin{array}{l}
N_{\rm L}^{(n)} \\
N_{\rm S}^{(n)}
\end{array}\right),
\end{equation}
where the maximum eigenvalue of the matrix is given by two.
This means that the sequence exhibits the self-similar structure
characterized by the rational number,
in contrast to the quasiperiodic sequence with the irrational one.
When a Thue-Morse sequence is assigned to the bonds
in the one-dimensional chain [see Fig.~\ref{fig: 1D2D}(a)],
with the corresponding lengths defined as $\ell$ and $s(\neq \ell)$,
the singular continuous pattern appears
in the lattice structure factor~\cite{Cheng_1988, Wolny_2000}.
This means that the sequence is not quasiperiodic, but
deterministically aperiodic.
It is important to note that the self-similar structure spatially appears when $\ell=s$.

Introducing the Thue-Morse sequence into the grid spacing of the triangular lattice,
we obtain the triangular lattice with Thue-Morse modulation.
The tiling structure with $\ell/s=2$ is shown in Fig.~\ref{fig: TMT}(a).
We clearly find that this tiling is composed of four kinds of triangles.
Figure~\ref{fig: TMT}(b) shows the modulated dice lattice
by the Thue-Morse sequence with $\ell=s$.
\cred{Here, we have defined two kinds of bonds as L and S bonds, respectively.
We have distinguished them,
introducing the open circle at the center of each L bond.
}
Comparing the tiling with the enlarged one by 4,
we find s substitution rule to generate the modulated dice lattice,
which is schematically shown in Fig.~\ref{fig: TMT}(c).
In the rule, there exist nine \cred{prototiles} which are distinguished
by their colors and hexagons inside of rhombuses.
The numbers of the tiles LR$_1$, LR$_2$, $\cdots$, SR$_3$ at iteration $n$ ${\bf v}^{(n)}=
(N_{\rm LR1}^{(n)}$, $N_{\rm LR2}^{(n)}$, $\cdots$ $N_{\rm SR3}^{(n)})^t$ satisfy
${\bf v}^{(n+1)}=M{\bf v}^{(n)}$ with
\begin{equation}
  M=\left(\begin{array}{lllllllll}
  2 & 0 & 0 & 0 & 0 & 0 &           0 & 6 & 2\\
  0 & 2 & 0 & 0 & 2 & 1 &           2 & 0 & 0\\
  2 & 0 & 3 & 0 & 3 & \frac{3}{2} & 2 & 6 & 6\\
  4 & 0 & 0 & 6 & 0 & 1 &           0 & 0 & 0\\
  0 & 0 & 0 & 0 & 6 & 1 &           4 & 0 & 2\\
  4 & 2 & 3 & 5 & 5 & 9 &           4 & 2 & 3\\
  0 & 6 & 4 & 0 & 0 & 0 &           2 & 0 & 0\\
  2 & 0 & 0 & 2 & 0 & 1 &           0 & 2 & 0\\
  2 & 6 & 6 & 3 & 0 & \frac{3}{2} & 2 & 0 & 3
  \end{array}\right).
\end{equation}
Since the maximum eigenvalue of the substitution matrix $M$ is $4^2$,
the tiling has the self-similar structure characterized by the rational number.

\begin{figure}[htb]
  \begin{center}
    \includegraphics[width=\linewidth]{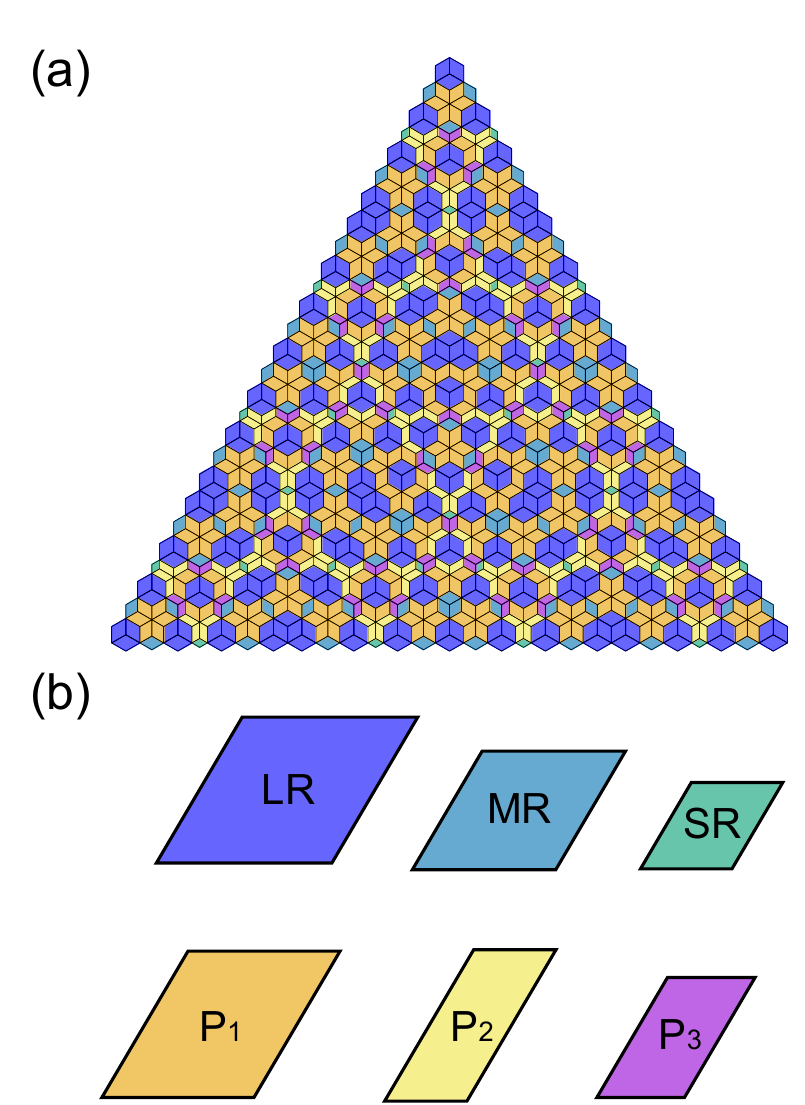}
    \caption{(a) Tribonacci-modulated dice lattice.
      (b) Large, medium and small rhombuses LR, MR and SM, and three parallelograms
      P$_1$, P$_2$ and P$_3$.
    }
    \label{fig: tri}
  \end{center}
\end{figure}

\begin{figure*}[htb]
  \begin{center}
    \includegraphics[width=\linewidth]{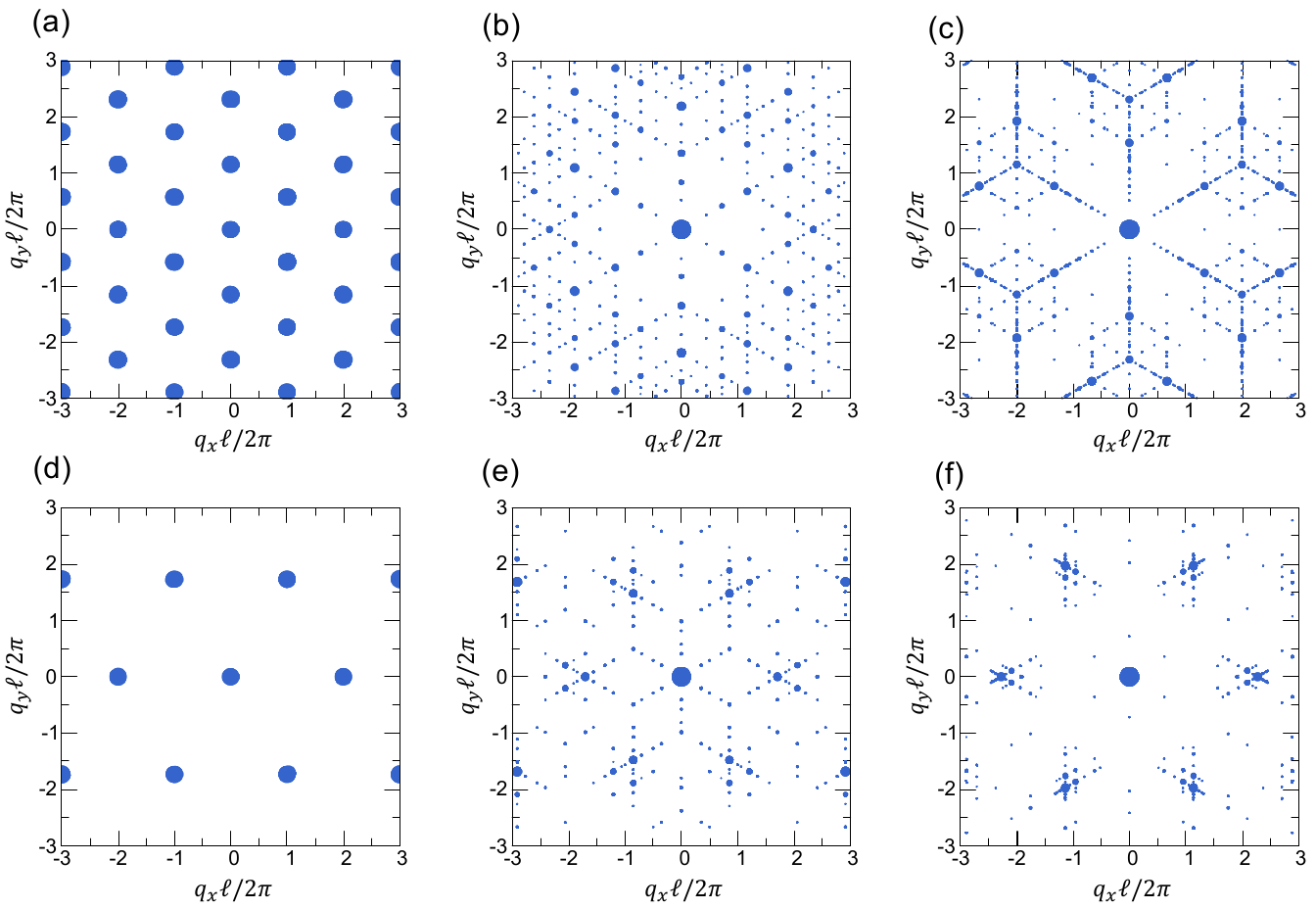}

\caption{Diffraction pattern.
(a) Periodic triangular lattice.
(b) Fibonacci-modulated triangular lattice.
(c) Thue-Morse modulated triangular lattice, showing singular continuous.
(d) Periodic dice lattice. 
(e) Silver-mean modulated dice lattice.
(f) Tribonacci-modulated dice lattice.
 The area of each circle represents the intensity at the position.
We cutoff the intensity with $S({\bm q}) < 0.0025$.
    }
    \label{fig: fourier}
  \end{center}
\end{figure*}

\subsection{Tribonacci tiling}
We consider the tribonacci sequence 
as an aperiodic one with multiple elements.
This sequence has the one-dimensional quasiperiodic structure
composed of three letters L, M, and S, as
\begin{equation}
{\rm LMLSLMLLMLSLMLMLSLMLLMLSLMLSLMLL} \cdots.\nonumber
\end{equation}
It is known that this can be generated
by a substitution rule:
L $\rightarrow$ LM, M $\rightarrow$ LS and S $\rightarrow$ L.
The total number of the letters
$N_{\rm tot}^{(n)}$ satisfies the recurrence formula
$N_{\rm tot}^{(n)} = N_{\rm tot}^{(n-1)} + N_{\rm tot}^{(n-2)} + N_{\rm tot}^{(n-3)}$,
where $N_{\rm tot}^{(n)}=\sum_i N_i^{(n)}$ and $N_i^{(n)}$ is the number of the letter $i$(=L, M, and S) at iteration $n$.
The numbers of the letters L, M and S at iteration $n$ satisfy
${\bf v}^{(n+1)}=M{\bf v}^{(n)}$ with
\begin{equation}
  M=\left(\begin{array}{lll}
1 & 1 & 1\\
1 & 0 & 0\\
0 & 1 & 0
  \end{array}\right),
\end{equation}
where ${\bf v}^{(n)}=\left(N_{\rm L}^{(n)}\; N_{\rm M}^{(n)}\; N_{\rm S}^{(n)}\right)^t$.
The maximum eigenvalue of the matrix $M$ is given by the tribonacci constant $\tau_T=(1+\kappa_++\kappa_-)/3\approx 1.839$
with $\kappa_\pm =(19\pm 3 \sqrt{33})^{1/3}$.
This means that the sequence has the self-similar structure
characterized by the irrational number.
When the tribonacci sequence is assigned to the bonds in one-dimensional chain
with three kinds of length scales $\ell, m,$ and $s$,
we obtain the modulated chain.
The self-similar structure spatially appears
when $s=\lambda, m=\lambda(\lambda+1)$, and $\ell=1$
with $\lambda=1/\tau_T$.
It is known that quasiperiodic behavior appears
in the structure factor~\cite{Baake_2020}.

The triangular lattice modulated by the tribonacci sequence should be complex
since other lengths in addition to three lengths $\ell$, $m$, and $s$ appear
and isosceles triangles have intricate angles.
Here, we demonstrate the modulated dice lattice
by a tribonacci sequence, as shown in Fig.~\ref{fig: tri}(a).
Since the kinds of letters is larger
than those of the generalized Fibonacci and Thue-Morse sequences,
the kinds of \cred{tiles} inherent in the tiling is also larger.
For example, this tribonacci tiling is composed of three rhobuses (LR, MR, SR)
and three parallelograms (P$_1$, P$_2$, P$_3$),
which are shown in Fig.~\ref{fig: tri}(b).
It is expected that
the substitution rule to generate the tribonacci tilings
should be obtained, comparing the tiling with the enlarged one by a certain irrational number.

\subsection{Diffraction pattern}
Finally,
we examine lattice structure factors $S({\bm q})$
to characterize the modulated triangular and dice lattices.
This quantity is defined by
\begin{eqnarray}
  S({\bm q})&=&\frac{1}{N}\left| \rho({\bm q}) \right|^2,\\
  \rho(\bm q) &=& \int \rho ({\bm r}) e^{-i {\bm q} \cdot {\bm r}} d {\bm r}=
  \sum_j e^{i {\bm q}\cdot {\bm r}_j},
\end{eqnarray}
where $\rho(\bm q) $ is the Fourier transform of the $\delta$-function scatterers
$\rho ({\bm r}) = \sum_j \delta({\bm r} - {\bm r}_j)$, and
${\bm r}_j$ is the coordinate of the $j$th vertex in the tiling,
and $N$ is the number of vertex sites.
The lattice structure factors reflect
the long-range orders in the tilings.
In fact, the periodic triangular and dice lattices have 
Bragg peaks with the six-fold rotational symmetry in the structure factors,
as shown in Figs.~\ref{fig: fourier}(a) and \ref{fig: fourier}(d).

The structure factors of the modulated triangular (dice) lattices
are shown in Figs.~\ref{fig: fourier}(b) and \ref{fig: fourier}(c)
[\ref{fig: fourier}(e) and \ref{fig: fourier}(f)].
We find that diffraction patterns have six-fold rotational symmetry.
Therefore, we confirm that these obtained aperiodic tilings are hexagonal.
We also find that the modulated lattices generated by the quasiperiodic sequences
have discrete Bragg peaks.
On the other hand, the Thue-Morse modulated triangular lattice 
with $\ell/s=2$ seems to have the singular continuous peak pattern, as shown in Fig.~\ref{fig: fourier}(c).

\begin{figure}[htb]
  \begin{center}
    \includegraphics[width=\linewidth]{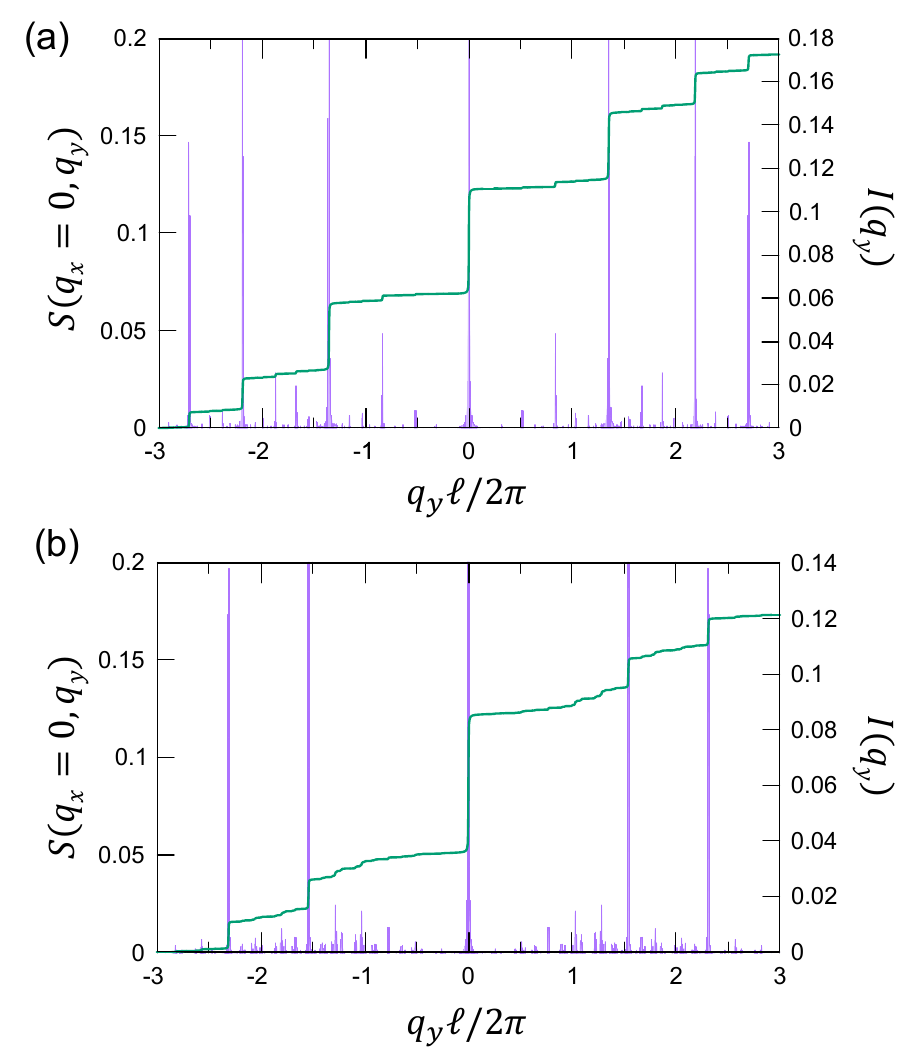}
    \caption{Diffraction patterns in details. (a) the golden-mean and (b) Thue-Morse tilings. 
    The purple and green lines represent the \cred{cross section} of lattice structure factor at $q_x=0$ and its integral, respectively.
    }
    \label{fig: fourier_1D}
  \end{center}
\end{figure}

The \cred{cross section} of $S({\bm q})$ is useful
to clarify the difference between the discrete and singular continuous peaks.
Figure~\ref{fig: fourier_1D} shows the \cred{cross section} along the $q_y$ axis and its integral $I(q_y)$
for the triangular lattices continuously modulated by
the Fibonacci and Thue-Morse sequences.
Here, $I(q_y)$ has been defined as
\begin{eqnarray}
  I(q_y) = \int^{q_y}_{q_0} S(q_x = 0, q^\prime_y)~dq^\prime_y,
\end{eqnarray}
with $q_0=-6\pi /\ell$.
We find in Fig.~\ref{fig: fourier_1D}(a) that 
the peaks are aperiodically located
and the integral $I(q_y)$ has jump singularities.
This is consistent with the fact that the tiling structure is quasiperiodic.
Figure~\ref{fig: fourier_1D}(b) shows a dense set of peaks in the structure factor of
the tiling based on the Thue-Morse sequence.
This leads to the smooth increase in $I(q_y)$,
which is in contrast to that for the quasiperiodic tilings.
By these results,
we confirm that the structure factors strongly reflect
the aperiodic sequences introduced into the grid spacing.

\section{Summary}\label{sec:summmary}
We have proposed a method to generate modulate triangular lattices using \cred{general aperiodic sequences}. In this method, aperiodic sequences are assigned to the spacings between the three grids of a regular triangular lattice. This process causes some grid intersections to split into three distinct intersections. The center of the three split intersections is defined as a single vertex. By combining these vertices with the remaining unsplit intersections, a modulated triangular lattice is constructed.
\cred{Furthermore, we have derived modulated dice (honeycomb) lattice by replacing the triangles with rhombuses and parallelograms (hexagons).}
We have demonstrated the generalized Fibonacci, Thue-Morse, and tribonacci tilings. Additionally, we have examined the lattice structure factor for various aperiodic tilings and observed that their peak patterns exhibit six-fold rotational symmetry, confirming that our tilings are hexagonal.
These results imply that our method provides a versatile framework for constructing a wide range of hexagonal aperiodic systems.

\begin{acknowledgments}
  The authors thank N.~Fujita for valuable discussions. Parts of the numerical calculations are performed in the supercomputing systems in ISSP, the University of Tokyo.
  T.M. was supported by JST SPRING, Japan Grant Numbers JPMJSP2106 and JPMJSP2180.
  This work was also supported by Grant-in-Aid for Scientific Research from JSPS,
  KAKENHI Grant Nos. 22K03525 (A.K.) and 24K06982 (T.D.).
\end{acknowledgments}

\appendix

\begin{figure}[tbh]
  \begin{center}
    \includegraphics[width=\linewidth]{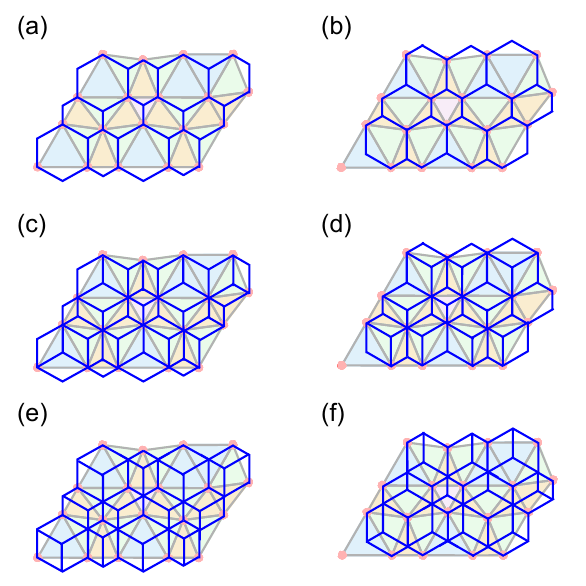}
    \caption{
      (a)[(b)] The replacement of upward (downward) triangles
      in the modulated triangular lattice with the hexagons.
      (c) and (e) [(d) and (f)]
      Two replacement rules of hexagons with rhombuses and parallelograms
      in the modulated honeycomb lattice shown in (a)[(b)].
    }
    \label{deco}
  \end{center}
\end{figure}

\begin{figure}[tbh]
  \begin{center}
    \includegraphics[width=\linewidth]{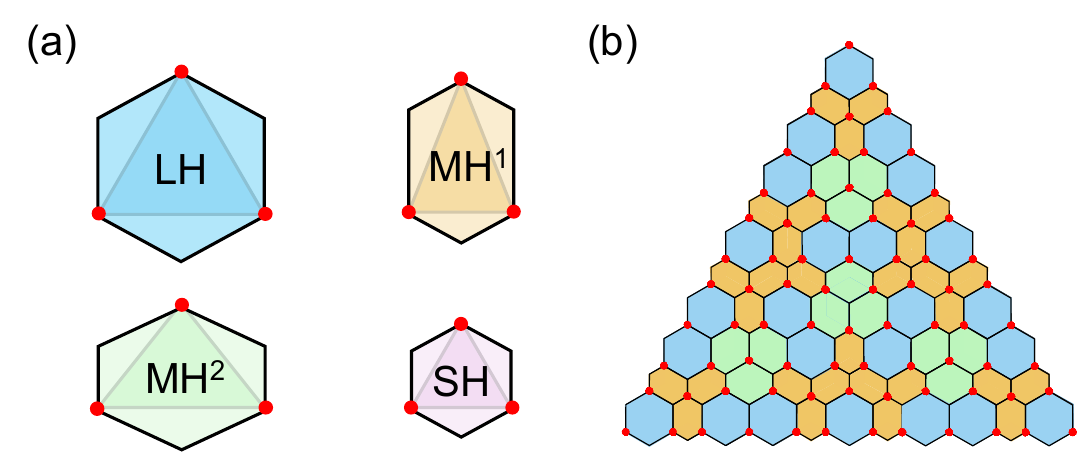}
    \caption{
      (a) The replacement of upward triangles,
      where LH, MH$^1$, MH$^2$, SH tiles are introduced instead of triangles.
      (b) Modulated honeycomb lattice
      using the rule (a).    
    }
    \label{fig: honey_up}
  \end{center}
\end{figure}

\begin{figure}[tbh]
  \begin{center}
    \includegraphics[width=\linewidth]{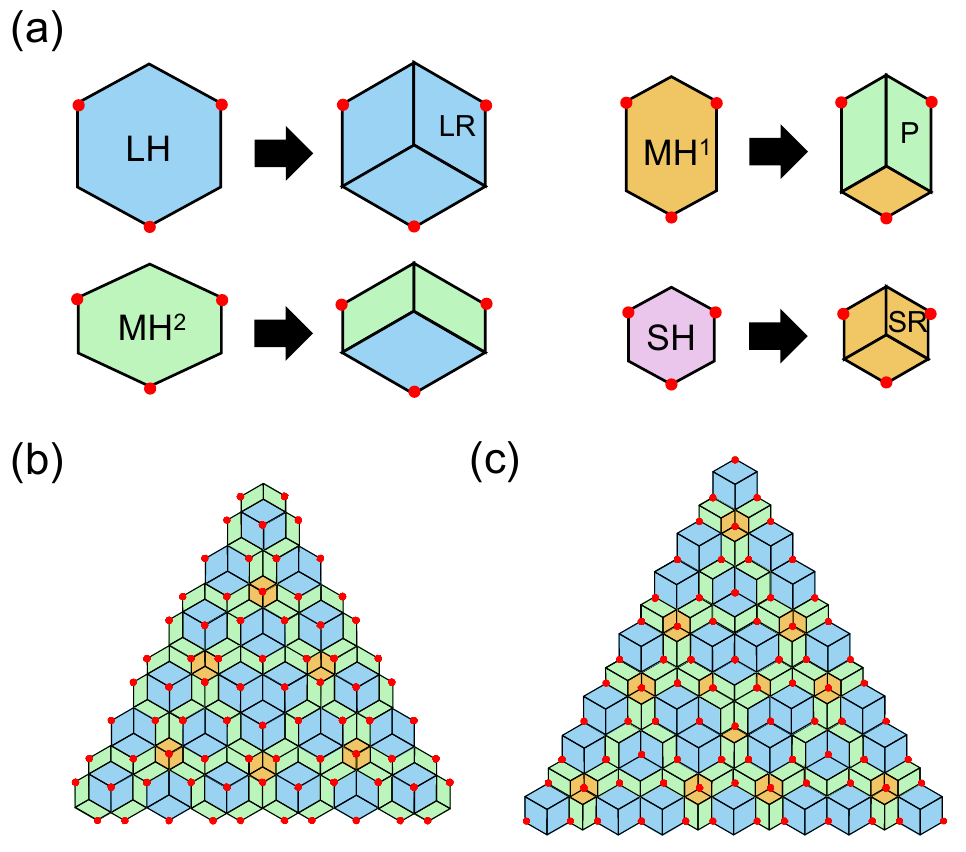}
    \caption{(a) A replacement of hexagons with LR, P, and SR tiles.
        (b)[(c)] Modulated dice lattice obtained from
        the modulated honeycomb lattices
        with the decoration rules of downward (upward) triangles.
    }
    \label{fig: dice2}
  \end{center}
\end{figure}

\begin{figure*}[tbh]
  \begin{center}
    \includegraphics[width=\linewidth]{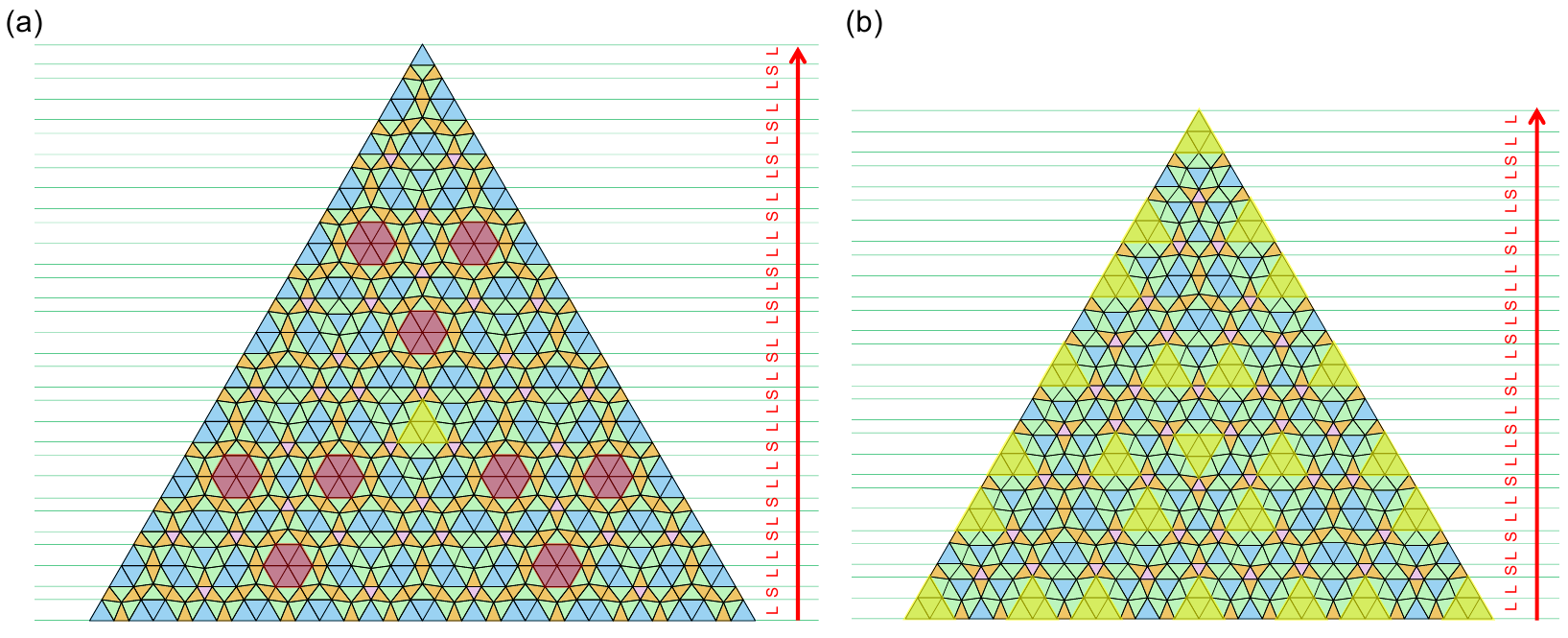}
    \caption{
    As an example, the vertically aligned sequence in the grids is illustrated.
      (a) A tiling generated by the Fibonacci sequence from the first letter. 
      (b) A tiling generated by the Fibonacci sequence from the third letter.
      The vertices composed of six large triangles are highlighted in red.
      The triangular clusters are highlighted in yellow.
      They are essentially different tilings belonging to different local isomorphism classes.
    }
    \label{fig: third}
  \end{center}
\end{figure*}

\begin{figure}[tbh]
  \begin{center}
    \includegraphics[width=0.9\linewidth]{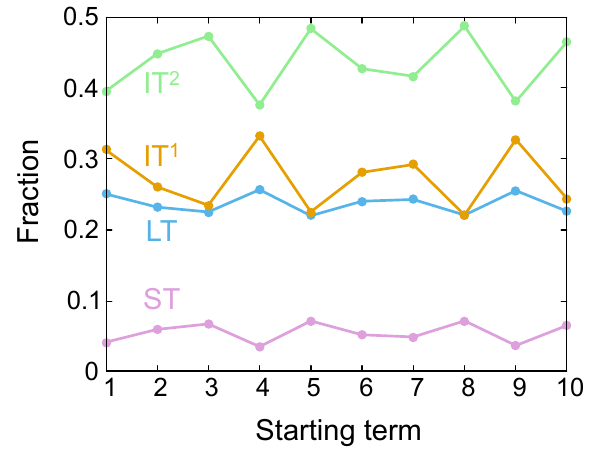}
    \caption{Tile fractions as a function of the starting term of the Fibonacci sequence
      in the modulated triangular lattices (see text).
      The results have been numerically computed by counting tiles in the tilings with $\approx 900\,000$ tiles.
    }
    \label{fig: fraction}
  \end{center}
\end{figure}

\section{A variety of continuously modulated dice and honeycomb lattices}\label{app}

In the main text, we have presented typical examples of
modulated dice and honeycomb lattices.
Here, we explain how to construct a variety of modulated dice and honeycomb lattices.

In the triangular lattice,
upward and downward triangles are arranged in an alternating pattern.
In the modulated triangular lattice, upward (downward) triangles can be replaced with
hexagons, as shown in Fig.~\ref{deco}(a) [Fig.~\ref{deco}(b)].
This replacement introduces LH, MH$^1$, MH$^2$, and SH tiles in the tiling,
determined by the positions of three nearest vertices.
Figure~\ref{fig: honey_up} shows the modulated honeycomb lattice
formed by replacing upward triangles,
while Figure~\ref{fig: honey_down} in the main text shows the alternative configuration.
The difference in the SH tile density indicates that these tilings are distinct.
Additionally, other modulated honeycomb lattice can be obtained
by considering the dual or Voronoi tessellation of the modulated triangular lattice.
This results in aperiodic tilings composed of the hexagons with complicated inner angles
(not shown).

We consider modulated dice lattices,
where hexagons in the simpler modulated honeycomb lattices
are replaced with rhombuses and parallelograms.
In the modulated honeycomb lattice shown in Fig.~\ref{deco}(a),
the hexagons are replaced with the LR, P, and SR tiles such that
their vertices with an acute angle {}\cred{60$^\circ$} (an obtuse angle \cred{120$^\circ$}) align with
the vertices of the modulated triangular lattice,
as shown in Fig.~\ref{deco}(c) [Fig.~\ref{deco}(e)].
The alternative honeycomb lattice shown in Fig.~\ref{deco}(b)
leads to the modulated dice lattice, as shown in Fig.~\ref{deco}(d) [Fig.~\ref{deco}(f)].
It is clear that two of the four modulated dice lattices,
which are shown in Figs.~\ref{deco}(c) and \ref{deco}(d), are identical,
as demonstrated in Fig.~\ref{fig: Dice} of the main text.
Therefore, three distinct types of modulated dice lattices are obtained.

As discussed in the main text, the arrangement of the Fibonacci sequence
introduces degrees of freedom.
Figures~\ref{fig: third}(a) and (b) show
the modulated triangular tilings generated from the Fibonacci sequences
starting with the first and third terms,
respectively.
A clear difference is observed in these tilings.
The former features hexagonal clusters, consisting of six adjacent LT tiles,
which are highlighted in red.
In contrast, no such clusters appear in the latter.
Moreover, both tilings have distinct densities
for the triangle clusters highlighted in yellow, consisting of four adjacent LT tiles.
Figure~\ref{fig: fraction} shows tile fractions as a function of the starting term of the Fibonacci sequence.
We could not find any peridodicity and regularity in tile fractions.
Therefore, these findings suggest that shifting the starting term in the sequence
leads to a wide variety of aperiodic tilings corresponding to the concept of local isomorphism classes~\cite{Levine_1986,Socolar_1986}.

\section{The aperiodic approximants of the honeycomb lattice}\label{sec: honey}
\begin{figure}[tbh]
  \begin{center}
    \includegraphics[width=0.65\linewidth]{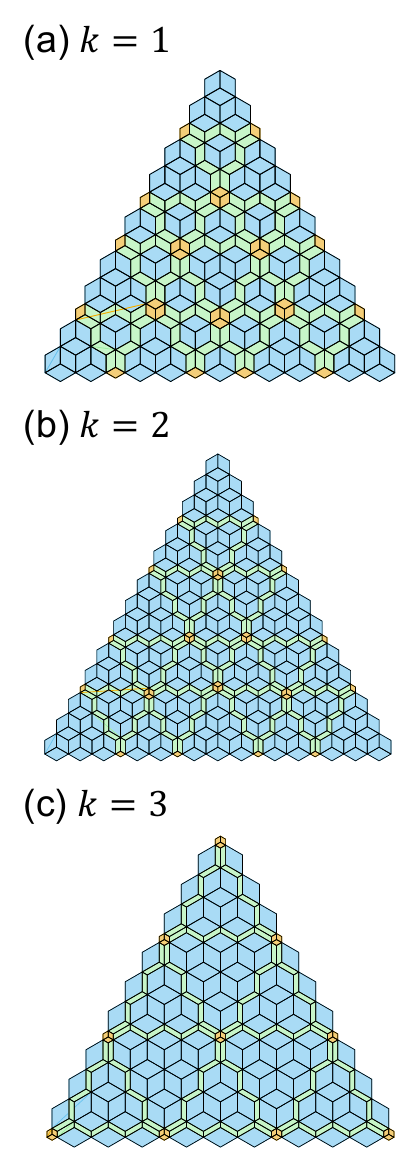}
    \caption{The aperiodic approximants of the honeycomb lattice reproduced through our method proposed in Sec.~\ref{sec: tiling}}.
    \label{fig: MMT}
  \end{center}
\end{figure}

Here, we demonstrate that our method generates aperiodic hexagonal tilings 
equivalent to those proposed previously~\cite{Sam_8GMT,Matsubara_honey}.
To this end,
we introduce the $k$th generalized Fibonacci sequence with an inversion center.
The substitution rule is given for $k=2m+1$ ($k=2m$): 
L $\rightarrow$ L$^\frac{1}{2}$L$^m$ S L$^m$L$^\frac{1}{2}$ (L $\rightarrow$ L$^m$ S L$^m$) and
S $\rightarrow$ L$^\frac{1}{2}$L$^\frac{1}{2}$
(S $\rightarrow$ L), where $m$ is integer.
The number of the letters L and S in the substitution rule is the same as 
that discussed in the main text,
indicating that the sequence is quasiperiodic with the metallic mean.
For convenience,
we define that, when L$^\frac{1}{2}$ appears at the ends of a sequence,
it is replaced with L.
We also consider two additional rules to generate symmetric sequences:
\begin{enumerate}
\item when the substitution rule is applied
to the leftmost and rightmost letters L on the sequence,
SL$^m$L$^\frac{1}{2}$ and L$^\frac{1}{2}$L$^m$S
(SL$^m$ and L$^m$S)
are generated for odd (even) $k$,
respectively.
\item when the deflation rule is applied to the leftmost and rightmost letters S on the sequence,
L$^\frac{1}{2}$L$^m$L$^\frac{1}{2}$ (L$^{m+1}$)
are generated for odd (even) $k$,
where $m = \left\lfloor \frac{k}{2} \right\rfloor$.
\end{enumerate}
The generated sequence is distinct from
the conventional Fibonacci sequence.
Introducing the symmetric generalized Fibonacci sequence 
into the line-spacing of three grids of the triangular lattice
and replacing the triangles with rhombuses and paralleograms,
we obtain the modulated dice lattices, as shown in Fig.~\ref{fig: MMT}.
This tiling is composed of large rhombuses with the edge length $\ell$, small rhombuses with the edge length $s$,
and parallelograms.
We find the large (small) regular hexagonal structure composed of three adjacent large (small) rhombuses.
An important point is that each rhombus always
belongs to a certain hexagonal region with edge length $\ell$ or $s$,
which is distinct from the tilings discussed in the main text.
This enables us to replace three adjacent rhombuses with large and small hexagons.
The aperiodic tilings obtained by these decorations are composed of
large and small hexagons, and parallelograms, and 
are equivalent to the hexagonal metallic-mean tilings proposed in Refs.~\cite{Sam_8GMT,Matsubara_honey}.
The $H_{\frac{1}{2} \frac{1}{2}}$ tiling can be also generated by our method
with further decorations (not shown).

\nocite{apsrev42Control}
\bibliographystyle{apsrev4-2}
\bibliography{./refs}

\end{document}